# Structural and magnetic properties of Ni doped $CeO_2$ nanoparticles

Shalendra Kumar*, Young. Joo. Kim, B. H. Koo, and C. G. Lee

*School of Nano & Advanced Materials Engineering, Changwon National University, 9 Sarim dong, Changwon-641-773, Korea*

*E-mail: shailuphy@gmail.com

**Abstract:** We report room temperature ferromagnetism in Ni doped $CeO_2$ nanoparticles using X-ray diffraction (XRD), high resolution transmission electron microscopy (HR-TEM), and dc magnetization measurements. Nanoparticles of $Ce_{1-x}Ni_xO_2$ ($0.0 \leq x \leq 0.10$) were prepared by using a co-precipitation method. XRD measurements indicate that all samples exhibit single phase nature with cubic structure and ruled out the presence of any secondary phase. Lattice parameters, strain and particle size calculated from XRD data have been found to decrease with increase in Ni doping. Inter-planner distance measured from HR-TEM images for different Ni doped samples indicate that Ni ions are substituting Ce ions in $CeO_2$ matrix. Magnetization measurements performed at room temperature display weak ferromagnetic behavior of $Ce_{1-x}Ni_xO_2$ ($0.0 \leq x \leq 0.10$) nanoparticles. Magnetic moment calculated from magnetic hysteresis loop was found to increases with Ni doping up to 7% and then start decreasing with further doping.
**Keywords:** $CeO_2$ Nanoparticles, XRD, HR-TEM, *DC* Magnetization

**Corresponding author**
E-mail: **shailuphy@gmail.com** (S. Kumar); **chglee@changwon.ac.kr** (C. G. Lee)
Ph: +82-55-213-3703; Fax: +82-55-261-7017

## 1. INTRODUCTION

During the last few years, dilute magnetic semiconductors (DMSs) have attracted the attention of the scientific community due to their advanced technological application in spintronics devices as well as to understand the underlying physics of these materials.[1,2] DMSs offer a possible system to realize the control of the charge transport by using the spin degree of freedom. DMSs can be designed and their properties can be tailored by doping of magnetic ion in a semiconducting host matrix. The original idea of doping the magnetic ion in a semiconducting host is to introduce interaction between magnetic atoms as a result, ferromagnetism (FM) can be induced in the semiconducting oxide host materials. However, in spite of progress in the search of new materials and device fabrication, a big issue has been discussed controversially at the moment is the nature of room temperature ferromagnetism (RTFM) in these systems. It is well known that much of the controversy has originated from the fact that clustering or phase separation of magnetic dopant ion can result in magnetic data that is misleading and unreliable.[3-5] Till date many potential compounds such as: pure and transition metal (TM) doped ZnO, $TiO_2$, $CeO_2$ and $HfO_2$ etc, have been observed to show RTFM but the intrinsic origination of the FM is still under contest.[6-11] Even thought the FM reported in the same system by different groups varies widely, ranging from RTFM to no magnetic ordering. Such a variation in magnetic properties indicates that FM in DMSs is very sensitive to preparation method and preparation conditions. Therefore, to take out the most precise information from magnetic data on a DMS materials, it is very important to perform a careful and accurate structural study.

In the present work, we have investigated the RTFM with different Ni concentration in $CeO_2$ nanoparticles. Structural properties studied using x-ray diffraction and high resolution transmission electron microscopy measurements clearly indicate that Ni doped $CeO_2$ particles with different Ni concentration have single phase cubic structure with nano-crystalline behavior. Magnetic properties infer that all the samples have week FM at RT and 7% Ni doped sample have the maximum value of the saturation magnetization.

## 2. EXPERIMENTAL DETAILS

Nanoparticles of Ni doped $CeO_2$ were synthesized using co-precipitation technique. All chemicals were dissolve in de-ionized water to get a 0.06 M solution with a molar ratio of x = [Ni]/(Ni+Ce). In this solution, $NH_4OH$ solution was added until the pH level reached 9. This mixture was stirred for 3 hours at room temperature and then washed 5 times with de-ionized water and alcohol. The precipitate obtained after wash was dried at 80 $^0C$ for 14 h. Finally, the dried samples were annealed at 500 $^0C$ for 4 hours to obtain $Ce_{1-x}Ni_xO_2$ nanoparticles. The prepared nanoparticles were characterized by the XRD, TEM and dc magnetization hysteresis loop measurements. Philips x-pert x-ray diffractometer with Cu $K\alpha$ ($\lambda$ = 1.54 Å) radiation was used to study single phase nature of the $Ce_{1-x}Ni_xO_2$ samples at room temperature. TEM measurements were performed using HR-TEM ((JEM 2100F). *DC* magnetization measurements were performed at 300K using Quantum Design physical properties measurement setup.

## 3. RESULTS AND DISCUSSION

Figure 1 shows the XRD pattern of the $Ce_{1-x}Ni_xO_2$ nanoparticles. XRD pattern indicates that all samples have single phase nature with cubic structure. It can be seen from Fig. 1 that none of the sample showed any extra peak corresponding to any impurity phase. The broadening of the diffraction peaks reflecting a small crystallite size and presence of the strain. The estimated particle size (not shown here) as a function of Ni doping is found to decrease from 9.2 to 5.3 nm. This infers that the substitution of the Ni obstructs the growth of $CeO_2$ nanoparticles which is similar to the earlier reported results in other TM doped metal oxides.[12] Lattice parameters calculated using Powder-X software, is found to decrease from 5.41 to 5.40Å with Ni doping. Earlier some groups[13, 14] have reported that for $CeO_2$ nanoparticles, the lattice parameters increases considerably with decrease in particle size which may be due to increase in the concentration of $Ce^{3+}$ ions replacing $Ce^{4+}$ ions as well as an increase in the number of oxygen vacancies. Although, in the present studied samples, we have observed that both particle size as well as lattice parameters decrease with increase in the doping of Ni in $CeO_2$ nanoparticles. This decrease in the lattice parameters may be due to difference in the ionic radii of Ni and Ce ions. It is observed that the pure and Ni doped $CeO_2$ nanoparticles are under considerable lattice strain which are in good agreement with the earlier reports on the $CeO_2$ nanoparticles. Recently, Thuber *et al.*[15] reported similar result in Ni doped $CeO_2$ nanoparticles. They observed that lattice strain decrease up to the 4% doping of Ni ions and found to increase with further doping of Ni ions. They explained that the decrease in the lattice strain is due to the replacement of Ce by Ni ions in the host matrix and the increase in the lattice strain above 4% doping is due to the interstitial incorporation of Ni ions. However, in the present case, we have observed that lattice strain decreases from $3.08 \times 10^{-3}$ to $1.96 \times 10^{-3}$ with Ni up to 10% doping. This observed behavior of particle size, lattice parameters and lattice strain indicates the all the Ni ions are replacing Ce ions in the $CeO_2$ host matrix.

Since XRD measurements cannot detect the presence of secondary phase at lower doping level due to their limitation. So, field emission transmission electron microscopy (FE-TEM) measurements were used to investigate the shape, size and presence of any secondary phase formation in $Ce_{1-x}Ni_xO_2$ ($0.0 \leq x \leq 0.10$) nanoparticles. Fig. 2(a-d) represents TEM images of $Ce_{1-x}Ni_xO_2$ ($0.0 \leq x \leq 0.10$) nanoparticles. It can be clearly seen from the TEM image that pure and doped $CeO_2$ particle are in nano-meter range and have almost spherical shape. In order to get more inside of the particle size distribution, we have measured the particle size using Image-J software. It is observed that the particle size decrease 8.5 to 3.5 nm with increase in the Ni concentration in $CeO_2$ host matrix. These results are in good agreement with the XRD results. Furthermore, to see any impurity phase in doped samples, we performed SAED and HR-TEM measurements of $Ce_{1-x}Ni_xO_2$ ($0.0 \leq x \leq 0.10$) nanoparticles. SAED patterns obtained by focusing the beam on the nanoparticles reveal the crystalline nature and phase purity of $Ce_{1-x}Ni_xO_2$ ($0.0 \leq x \leq 0.10$) nanoparticles. SAED pattern demonstrate that each nanoparticle is indeed in single phase. Moreover, HR-TEM images were also taken at different part of the sample to investigate the presence of any secondary or NiO phase in $Ce_{1-x}Ni_xO_2$ ($0.0 \leq x \leq 0.10$) nanoparticles. Fig. 3(a-d) shows the HR-TEM image of Ni doped $CeO_2$ nanoparticles. A careful analysis of the inter-planer distance calculated from HR-TEM image shows the planes of cubic $CeO_2$. HR-TEM image taken on the samples with different Ni concentration also indicate that individual nanoparticle is a single crystal. HR-TEM observations of large number of randomly selected particles showed the absence of impurity phase and suggesting their good crystallinity.

RTFM of Ni doped $CeO_2$ has been studied using magnetic hysteresis loop measurements. In Fig. 4, we have shown typical magnetization curve *(M)* versus field curves of $Ce_{1-x}Ni_xO_2$ ($0.0 \leq x \leq 0.10$) nanoparticles measured at RT. The observed value of the magnetic moment clearly indicates the week ferromagnetic behavior Ni doped $CeO_2$ nanoparticles. The coercivity ($H_C$) field values calculated from hysteresis was in the range from 20-85 Oe for different Ni concentration. It is exciting to see that even pure $CeO_2$ nanoparticles display the ferromagnetic behavior (see inset in Fig. 4) at room temperature with coercivity of ~ 20 Oe. The result is pretty interesting because bulk $CeO_2$ is an insulator with $Ce^{4+}$ in the $4f^0$ configuration. However, similar type of results is also reported by some groups.[21, 22] According to Sundaresan *et al.*[13] the weak ferromagnetism in pure $CeO_2$ as well as in other undoped oxide semiconductors results from the exchange interactions between electron spin moment resulting from the oxygen vacancies at the particle surface. Ge *et al.*[14] also reported the FM in pure $CeO_2$ nano cubes. They reported that FM in pure $CeO_2$ is originated from oxygen vacancies because oxygen vacancies cause spin polarization of *f* electron of Ce ions surrounding oxygen vacancies. Moreover, in the present study, we have observed that magnetization of $Ce_{1-x}Ni_xO_2$ increase with Ni concentration, and after x=0.07, the magnetization starts decreasing.

The observed FM behavior of $Ce_{1-x}Ni_xO_2$ nanoparticle has been explained in the light of F-centre exchange coupling (FCE). In FCE mechanism, an electron trapped in oxygen vacancy creates F-centre and occupies an orbital which overlaps the *d* shell of neighboring Ni ions. According to the Hund's rule and Pauli Exclusion Principle, the trapped electrons spin should have direction parallel to two neighboring Ni ions, which results ferromagnetic ordering. It is well known that $CeO_2$ is often used as storage medium because oxygen vacancy can be easily formed in $CeO_2$. Moreover, the doping of Ni ions may enhance the number of oxygen vacancy in doped $CeO_2$ to maintain charge neutrality in the system. Generally, magnetic ordering in insulating oxides is due to the super-exchange coupling which results antiferromagnetic ordering. The observed RTFM in Ni doped $CeO_2$ result from the competition between the FCE coupling and super-exchange coupling. It is observed that saturation magnetization increases with increase in the Ni contents in $CeO_2$ (up to x=0.07) and then start decreasing with further increase in the Ni ion concentrations. This complex magnetic behavior can be explained as follow: When the concentration of Ni ions is increased beyond a percolation limit (in the present case x= 0.07), they are mediated my FCE interactions due the small separation between Ni ions, which result increase of magnetization with Ni contents. However, further increase in Ni ions may results super-exchange interactions between Ni ions mediated by oxygen ions. This super-exchange interaction results the antiferromagnetic ordering, which reduce the FM ordering in the Ni doped $CeO_2$.

## 4. CONCLUSIONS

In summary, we have successfully prepared the nanoparticles of Ni doped $CeO_2$ using co-precipitation technique. XRD HR-TEM and SAED results indicate that all the samples have single phase polycrystalline nature. DC magnetization measurements reveal that all the samples exhibit weak RT-FM.

**Figure Captions**

**Fig. 1.** (Colour Online) X-ray diffraction patterns of $Ce_{1-x}Ni_xO_2$ ($0.0 \leq x \leq 0.10$) nanoparticles.
**Fig. 2.** TEM micrograph and corresponding SAED patters of $Ce_{1-x}Ni_xO_2$ ($0.0 \leq x \leq 0.10$) nanoparticles (a) *x* = 0.0, (b) *x* = 0.01, (c) *x* = 0.07, and (d) *x* = 0.10.
**Fig. 3.** (Colour Online) HR-TEM image of $Ce_{1-x}Ni_xO_2$ ($0.0 \leq x \leq 0.10$) nanoparticles (a) *x* = 0.0, (b) *x* = 0.01, (c) *x* = 0.07, and (d) *x* = 0.10. Inset shows the zoom part of HR image of the selected area.
**Fig. 4.** (Colour Online) Room temperature magnetization data of $Ce_{1-x}Ni_xO_2$ ($0.0 \leq x \leq 0.10$) nanoparticles.

# Structural and magnetic properties of Ni doped CeO$_2$ nanoparticles

Shalendra Kumar*, Young. Joo. Kim, B. H. Koo, and C. G. Lee

*School of Nano & Advanced Materials Engineering, Changwon National University, 9 Sarim dong, Changwon-641-773, Korea*

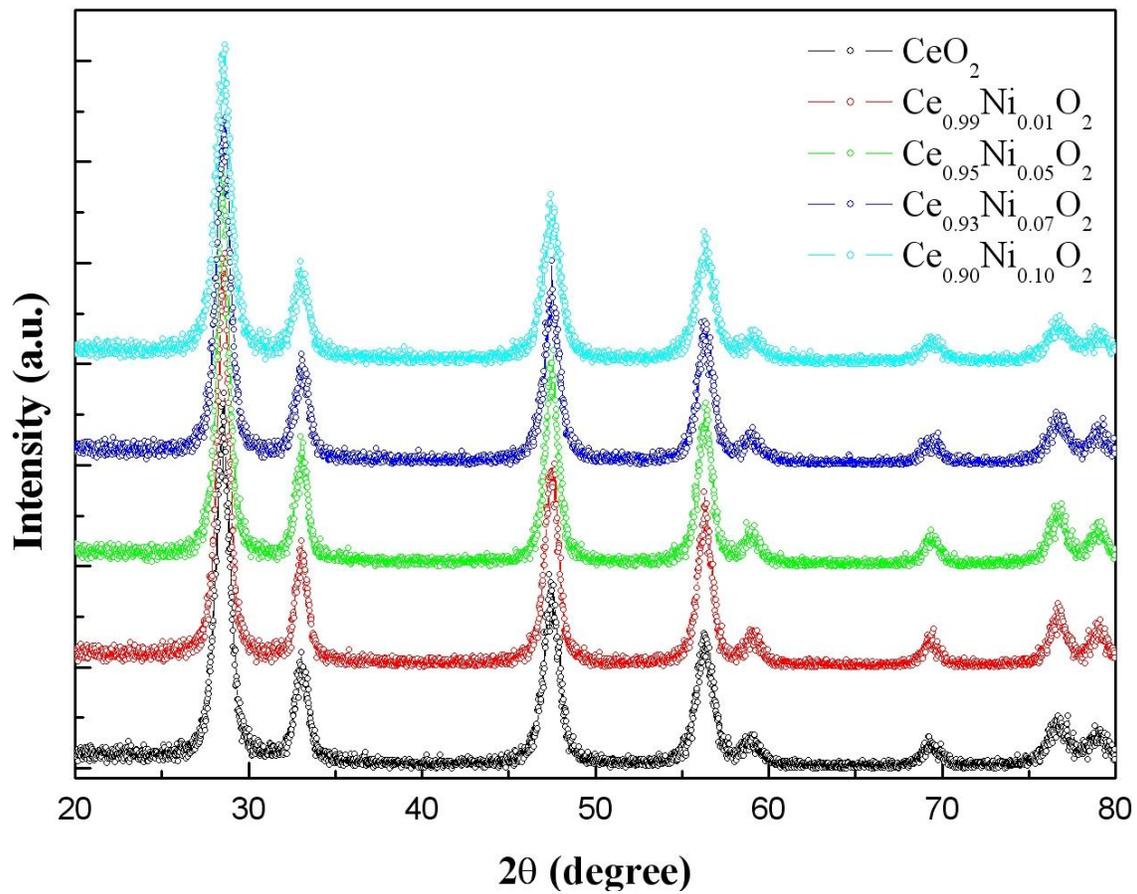

**Fig. 1**

# Structural and magnetic properties of Ni doped CeO₂ nanoparticles

Shalendra Kumar*, Young. Joo. Kim, B. H. Koo, and C. G. Lee

*School of Nano & Advanced Materials Engineering, Changwon National University, 9 Sarim dong, Changwon-641-773, Korea*

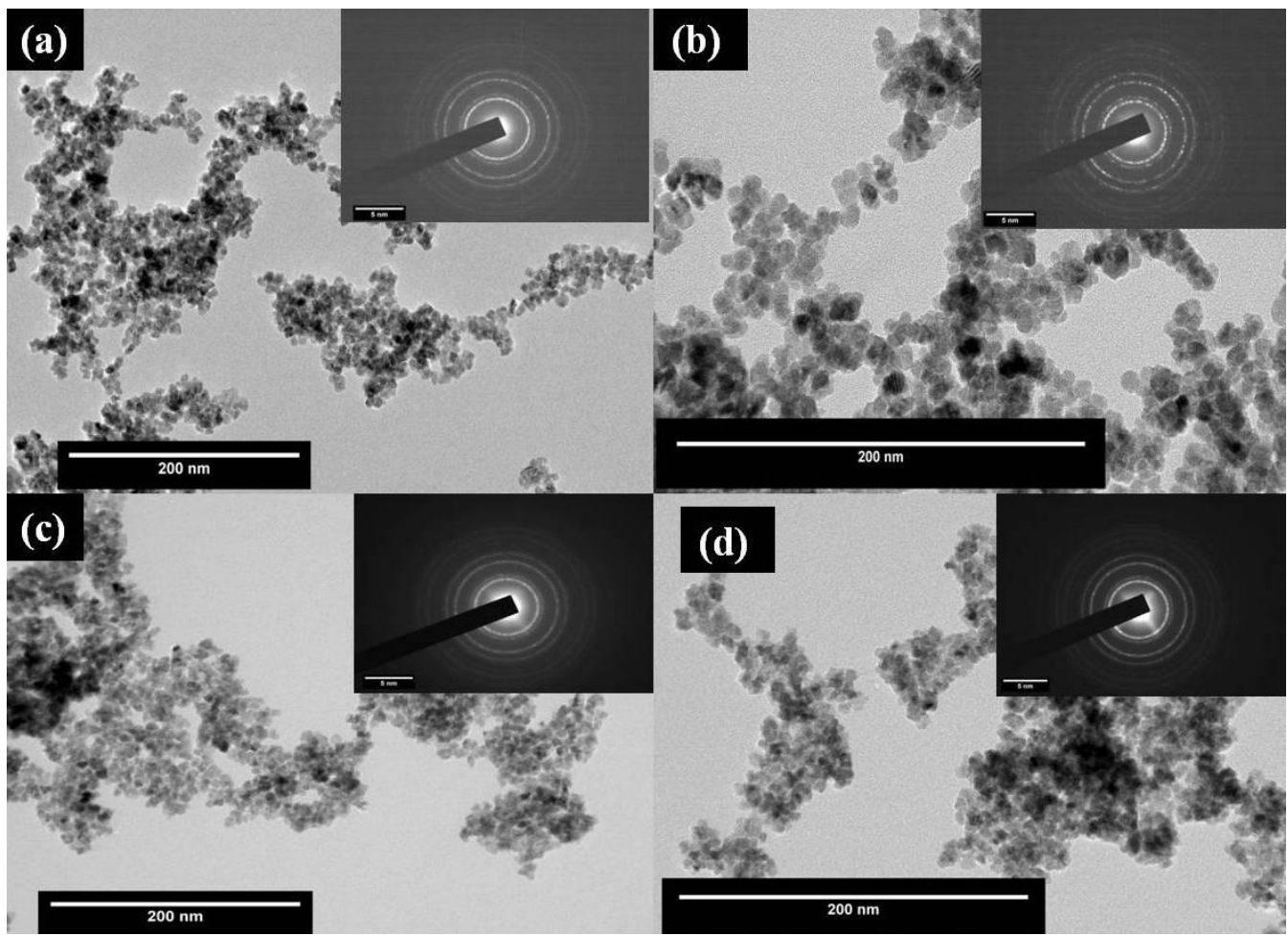

**Fig. 2**

# Structural and magnetic properties of Ni doped CeO₂ nanoparticles

Shalendra Kumar*, Young. Joo. Kim, B. H. Koo, and C. G. Lee

*School of Nano & Advanced Materials Engineering, Changwon National University, 9 Sarim dong, Changwon-641-773, Korea*

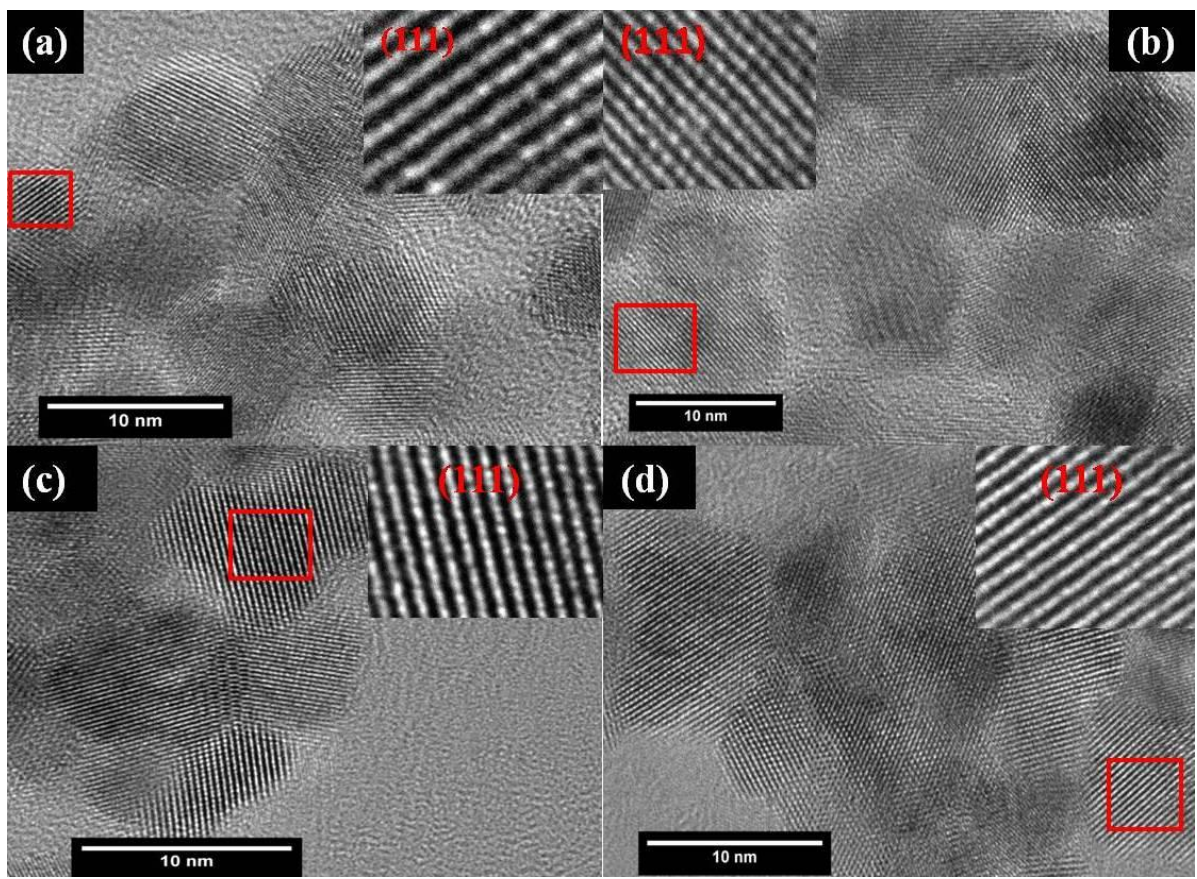

**Fig. 3**

# Structural and magnetic properties of Ni doped CeO₂ nanoparticles

Shalendra Kumar*, Young. Joo. Kim, B. H. Koo, and C. G. Lee

*School of Nano & Advanced Materials Engineering, Changwon National University, 9 Sarim dong, Changwon-641-773, Korea*

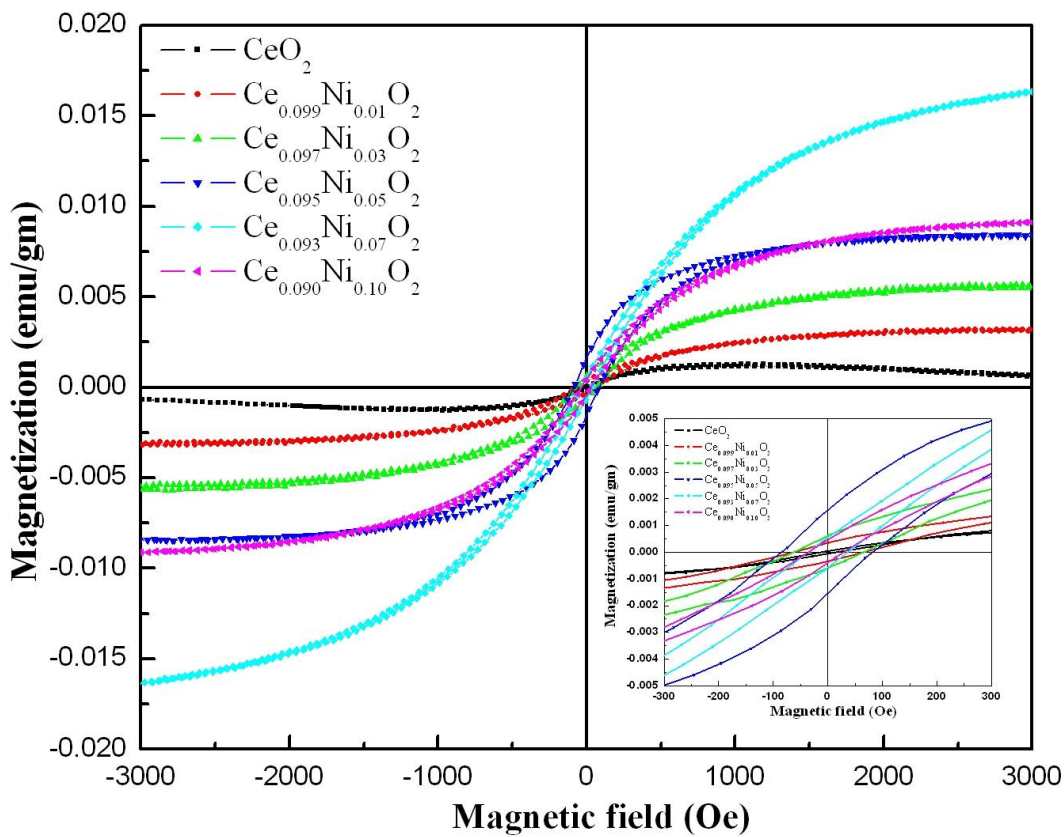

**Fig. 4**